\documentstyle[aps,amssymb,epsf,twocolumn,eepic]{revtex}
\begin{document}
\title{Using entanglement improves precision of quantum measurements}
\author{G. Mauro D'Ariano$^{a,b,c}$\cite{pa}, Paoloplacido Lo Presti$^a$, Matteo G. A. Paris$^a$} 
\address{$^a$ Quantum Optics \& Information Group\cite{ems,urls}, Istituto Nazionale di
Fisica della Materia, Unit\`a di Pavia \\
$^b$ Istituto Nazionale di Fisica Nucleare, Sezione di Pavia \\
Dipartimento di Fisica ``A. Volta'', via Bassi 6, I-27100 Pavia, Italy\\
$^c$ Department of Electrical and Computer Engineering, Northwestern
University, Evanston, IL  60208} 
\date{\today}
\maketitle
%%%%%%%%%%%%%%%%%%%%%%%%%%%%%%%%%%%%%%%%%%%%%%%%%%%%%%%%%%%%%%%
\begin{abstract}
We show how entanglement can be used to improve the estimation of an
unknown transformation. Using entanglement is always of benefit, in
improving either the precision or the stability of the
measurement. Examples relevant for applications are illustrated, for
either qubits and continuous variables.
\end{abstract}
%%%%%%%%%%%%%%%%%%%%%%%%%%%%%%%%%%%%%%%%%%%%%%%%%%%%%%%%%%%%%%%%%%%%%%%%%%%%
\date{\today}
\maketitle
\pacs{PACS numbers:  03.65.Wj}
Entanglement is certainly the most distinctive feature of quantum
mechanics.  The quantum nonlocality due to entanglement, which has
puzzled generations of theoreticians since the work of Einstein,
Podolsky, and Rosen\cite{epr}, in the last decade eventually has been
harnessed for practical use in the new quantum information
technology\cite{popescubook,nielsen}.  Entanglement has become
the essential resource for quantum computing, quantum teleportation,
and secure cryptographic protocols\cite{nielsen}. Recently,
entanglement has been proved as a valuable resource for improving 
optical resolution\cite{fabre}, 
spectroscopy\cite{spettr}, quantum lithography\cite{qlit}, and has
shown to be a crucial ingredient for making the tomography of a
quantum device\cite{tomochannel}, with a single input entangled state
playing the role of all possible states at the input of the
device---another manifestation of the {\em quantum parallelism}, the 
feature of entanglement that is the core of quantum computing
algorithms \cite{shor,grover}.

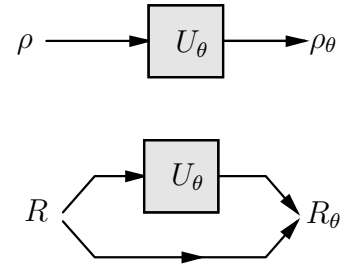
\begin{figure}[hbt]
%\vskip .5truecm
\begin{center}
\setlength{\unitlength}{0.004mm}
\begin{picture}(10393,8673)(0,-10)
\thicklines
\texture{0 0 0 888888 88000000 0 0 80808 8000000 0 0 888888 88000000 0 0 80808 
	8000000 0 0 888888 88000000 0 0 80808 8000000 0 0 888888 88000000 0 0 80808 }
\path(6847,7404)(9572,7404)
\path(6847,7404)(9572,7404)
\blacken\path(8972.000,7254.000)(9572.000,7404.000)(8972.000,7554.000)(8972.000,7254.000)
\path(972,7404)(4372,7404)
\blacken\path(3772.000,7254.000)(4372.000,7404.000)(3772.000,7554.000)(3772.000,7254.000)
\path(6697,2904)(8272,2904)(9322,1704)
\blacken\path(8814.011,2056.770)(9322.000,1704.000)(9039.784,2254.322)(8814.011,2056.770)
\path(1522,1704)(2572,2904)(4222,2904)
\blacken\path(3622.000,2754.000)(4222.000,2904.000)(3622.000,3054.000)(3622.000,2754.000)
\path(1522,1404)(2572,204)(8272,204)(9322,1404)
\blacken\path(9039.784,853.678)(9322.000,1404.000)(8814.011,1051.230)(9039.784,853.678)
\path(5872,204)(6097,204)
\blacken\path(5497.000,54.000)(6097.000,204.000)(5497.000,354.000)(5497.000,54.000)
\texture{0 0 0 888888 88000000 0 0 80808 8000000 0 0 888888 88000000 0 0 80808 
	8000000 0 0 888888 88000000 0 0 80808 8000000 0 0 888888 88000000 0 0 80808 }
\shade\path(4372,8604)(6847,8604)(6847,6204)
	(4372,6204)(4372,8604)
\path(4372,8604)(6847,8604)(6847,6204)
	(4372,6204)(4372,8604)
\shade\path(4222,4104)(6697,4104)(6697,1704)
	(4222,1704)(4222,4104)
\path(4222,4104)(6697,4104)(6697,1704)
	(4222,1704)(4222,4104)
\put(5272,7029){\makebox(0,0)[lb]{\smash{\large $U_{\theta}$}}}
\put(247,7179){\makebox(0,0)[b]{\smash{\large $\rho$}}}
\put(10200,7179){\makebox(0,0)[b]{\smash{\large $\rho_{\theta}$}}}
\put(5122,2604){\makebox(0,0)[lb]{\smash{\large $U_{\theta}$}}}
\put(622,1404){\makebox(0,0)[b]{\smash{\large $R$}}}
\put(10200,1329){\makebox(0,0)[b]{\smash{\large $R_{\theta}$}}}
\end{picture}
\end{center}
\caption{Measurement schemes considered in the present letter. The
parameter $\theta$ of the density operator $\rho_{\theta}$ is
estimated as the result of a unitary transformation 
$\rho\to\rho_{\theta}=  U_{\theta}\rho U^\dag_{\theta}$ (up
figure).  In this scenario the use of an entangled input
$R$ in place of $\rho$ is considered, with the unknown
transformation $U_{\theta}$ acting locally on one Hilbert space only (down
figure).} \label{f-scheme}\end{figure}

In this letter we will show how in general entanglement can be
used to improve quantum measurements, for either precision or
stability. The measurement scheme will be considered in the general
framework of quantum estimation theory\cite{helstrom}, in which one
needs to estimate the parameter $\theta$ of the density operator
$\rho_{\theta}$ on the Hilbert space ${\cal H}$ as the result of a
unitary  transformation $\rho\to\rho_{\theta}= 
U_{\theta}\rho U^\dag_{\theta}$---more generally a quantum operation
${\mathrm Q}_{\theta}$ could be considered, with
$\rho_{\theta}={\mathrm Q}_{\theta}(\rho)$,  corresponding to  a
parameter of any physical (amplifying, measuring, etc.) device. This
situation for known input state $\rho$  is very common in practice,
e. g. in interferometry\cite{shapiro}, and more generally whenever the 
measurement is {\em indirect}, resorting to the detection of a
change in an ancillary part of the measuring apparatus.
In this scenario we will consider the use of an entangled input state
$R$ in place of $\rho$,  with the unknown transformation $U_{\theta}$
acting locally only on one side of the entangled state.  In
tensor notation: $R\to R_{\theta}=U_{\theta}\otimes I\, R\,
U^\dag_{\theta}\otimes I$. The situation is depicted in Fig.
\ref{f-scheme}. As we will see in this letter, the entangled
configuration is better than the conventional one, for either precision
or stability of the measurement. This is due to the fact that, in some
sense, the input entangled state is equivalent to many input states in
``quantum parallel''. In the following we will examine  different
measurement situations separately, and we will draw general
conclusions at the end.

\medskip
{\em Covariant measurements}.---In a covariant measurement the
parameter $\theta$ is the element 
$g\in{\mathbf G}$  of a group ${\mathbf G}$ of transformations. This
kind of measurement has been thoroughly analyzed in Ref. \cite{holevo}.

Let us first illustrate the mechanism of entanglement on a
simple example. We want to discriminate among the four
unitary transformations represented by the Pauli matrices 
$\sigma_0\equiv I\,,\sigma_1\equiv\sigma_x\,,
\sigma_2\equiv\sigma_y\,,\sigma_3\equiv\sigma_z$. As well
known, they form a unitary discrete group \cite{D2}.
By applying the four transformations to any single-qubit input state
$|\psi\rangle\in{\mathbf C}^2$ we always obtain four linearly
dependent states, which makes the conventional scheme in
Fig. \ref{f-scheme} useless for a reliable discrimination. On the
contrary, if we apply the four 
matrices to the maximally entangled input state 
${1\over\sqrt{2}}|I\rangle\!\rangle$ we obtain the four Bell states
$\sigma_j\otimes I {1\over\sqrt{2}}|I\rangle\!\rangle\equiv
{1\over\sqrt{2}}|\sigma_j\rangle\!\rangle$,  which are
mutually orthogonal. Here we use the notation
$|A\rangle\!\rangle\doteq\sum_{ij}A_{ij}|i\rangle|j\rangle\equiv
A\otimes |I\rangle\!\rangle$, which puts vectors $|A\rangle\!\rangle\in
{\cal H}\otimes{\cal H}$ into correspondence with operators $A$ on
${\cal H}$, $A_{ij}$ denoting the matrix elements of $A$
on the fixed basis $\{|i\rangle\}$ for ${\cal H}$,  and $I$ being
the identity operator. This simple example is very instructive: the
discrimination among the four Pauli transformations $\sigma_j$, which is
impossible with a single qubit input state, becomes possible
and exact when applying $\sigma_j$ to a maximally
entangled state. The mechanism is clear: using an entangled state
instead of a single qubit, doubles the dimension of the Hilbert
space ${\cal H}_{out}$ spanned by the output states, allowing perfect
discrimination of the four $\sigma_j$. This example can be
generalized easily to any dimension $d$, when  discriminating among
the $d^2$ unitary transformations $U(m,n)=\sum_{k=0}^{d-1} e^{2\pi
ikm/d}|k\rangle\langle k\oplus n|$, $n$ and $m$ ranging in $0\div d-1$, and $\oplus$
denoting addition modulo $d$\cite{Zd}. Now, using the 
maximally entagled state ${1\over\sqrt{d}}|I\rangle\!\rangle$ at the input will
produce the $d^2$ orthogonal output states $U(m,n)\otimes
{1\over\sqrt{d}}|I\rangle\!\rangle$, which allows perfect 
discrimination among all $U(m,n)$, whereas a non entangled input 
$|\psi\rangle\in{\cal H}$ would output $d^2$ linearly dependent states in 
the $d$-dimensional ${\cal H}$. More generally, let us
consider a set of unitary transformations $\{U_g\}$, $g\in{\mathbf G}$
that form a (projective) representation of the group ${\mathbf G}$,
i. e. $U_gU_h=\omega(g,h) U_{gh}$, where  $\omega(g,h)$ is a suitable
phase\cite{cocycle}. For simplicity let us consider the case of an
irreducible representation (the reducible case is technically more
complicate, and needs the knowledge of all irreducible components on
invariant subspaces). For every operator $O$ on ${\cal H}$, from the
Schur's lemma one has the trace identity 
\begin{eqnarray}
[U_gOU^\dag_g]_{\mathbf G}=\mbox{Tr}[O]\;,
\label{tri}
\end{eqnarray}
where $[f(g)]_{\mathbf G}$ denotes the group averaging
$[f(g)]_{\mathbf G}\doteq\frac{d}{|\mathbf G|} \sum_{g\in\mathbf
G}f(g)$ with suitable normalization, $|{\mathbf G}|$ the 
cardinality of ${\mathbf G}$. Eq. (\ref{tri}) 
generalizes to the continuous case for group averaging
$[f(g)]_{\mathbf G}\doteq\int_{\mathbf G}
\mbox{d}g  f(g)$,  $\mbox{d}g $ being a (normalized)
invariant measure on ${\mathbf G}$. For a general input state
$|E\rangle\!\rangle\in{\cal H}\otimes{\cal H}$, the Hilbert space
${\cal H}_{out}$ spanned by the output states is the 
support of the operator  $O=[|\Psi_g\rangle\!\rangle\langle\!\langle \Psi_g|]_{\mathbf
G}$, with  $\Psi_g=U_gE$. One has $O=\mbox{Tr}_1 [|E\rangle\!\rangle\langle\!\langle
E|]\otimes I=(E^\dag E)^{T}\otimes I$, $\mbox{Tr}_1$
representing the partial trace over the first Hilbert 
space, and $T$  denoting transposition with respect to the basis
$\{|i\rangle\}$ for ${\cal H}$. Therefore, dim$({\cal 
H}_{out})=d\times$rank$(E)$, and since rank$(E)$ is equal to the
Schmidt number of $|E\rangle\!\rangle$, we conclude that an entangled
input always increases the dimension of ${\cal H}_{out}$, i. e. 
it improves the precision of the measurement.
\par Since the Schmidt number does not depend on the 
actual amount of entanglement of $|E\rangle\!\rangle$, a
more refined goodness criterion can be 
given in terms of the Holevo bound\cite{nielsen} 
$\chi=S([|\Psi_g\rangle\!\rangle\langle\!\langle\Psi_g|]_{\mathbf G})-
[S(|\Psi_g\rangle\!\rangle\langle\!\langle\Psi_g|)]_{\mathbf G}$ for the
information accessible from the measurement, $S$ denoting the von
Neumann entropy. Eq. (\ref{tri})  gives $\chi=d^{-1}\log
d+S[E^{\dag}E]$,  i. e. the bound is increased exactly of the amount
of entanglement  $S[E^{\dag}E]$ of the input state. 
\par With the measurement problem addressed in a maximum likelihood
strategy, it is easy to see that the optimal POVM $\mbox{d}\Pi_g$ is
of the form  
\begin{eqnarray}
\mbox{d}\Pi_g=\mbox{d}gU_g\otimes I SU_g^\dag\otimes I\;,
\end{eqnarray}
with $S\ge 0$ a positive operator on ${\cal H}\otimes{\cal H}$
normalized as $\mbox{Tr}_1[S]=I$. By covariance, the
maximum average likelihood is equal to $\langle\!\langle
E|S|E\rangle\!\rangle\leq d$, since normalization limits the maximum
eigenvalue of S below $d$. The bound is saturated for $E=d^{-1/2}U$, with
$U$ unitary, i. e. for maximally entangled input. 
\par Another way to see the optimality of a maximally entangled input
is to notice that the average overlap
$\Omega(E)\doteq\left[|\langle\!\langle\Psi_g|E\rangle\!\rangle|^2\right]_{\mathbf
G})\equiv\mbox{Tr}[(E^\dag E)^2)]$ is a Schur convex
function of the reduced density operator $E^\dag E$.
Following Ref. \cite{nielsenprl}, this implies that if
$|A\rangle\!\rangle\prec|B\rangle\!\rangle$  ($|A\rangle\!\rangle$ is
``majorized'' by $|B\rangle\!\rangle$), 
then $\Omega(A)\leq\Omega(B)$, whence the minimum averaged overlap
between output states comes from a maximally entangled input,
since this is majorized by any other state. That the optimal
estimation strategy to discriminate among unitaries needs 
entangled inputs has also been noticed in Ref. \cite{vidal}. 
\par As an example in infinite dimensions, consider 
the problem of estimating the displacement of a harmonic oscillator
in the phase space, i. e. the parameter $\alpha\in{\mathbb C}$ of the
transformation $\rho\to\rho_\alpha =D(\alpha)\rho D^\dag(\alpha)$,
where $D(\alpha)=\exp(\alpha a^\dag-\overline{\alpha}a)$ is the
displacement operator for annihilation and creation operators $a$ and
$a^\dag$ respectively (in this case ${\mathbf G}$ is the 
Weyl-Heisenberg group). For unentangled $\rho$, an estimation of $\alpha$ 
isotropic on ${\mathbb C}$ is equivalent to a
optimal joint measurement of position and momentum, which, 
as well known, is affected by a unavoidable minimum noise of
3dB\cite{goodman}. Here, the optimal state (for fixed minimum energy)
is the vacuum, and the corresponding conditional probability of
measuring $z$ given $\alpha$ is
$p(z|\alpha)=\pi^{-1}\exp[-|z-\alpha|^2]$. Now, consider the case in
which the estimation is made with $D(\alpha)$ acting on the
entangled state
\begin{eqnarray}
|E\rangle\!\rangle=\sqrt{1-|x|^2}\sum_{n=0}^{\infty}x^n|n\rangle|n\rangle\;,\label{dw}
\end{eqnarray}
with $|x|\le 1$ (the state (\ref{dw}) can be achieved by parametric
downconversion of vacuum). Here, we can use the orthonormal resolution
of the identity $|D(z)\rangle\!\rangle\langle\!\langle D(z)|$ of
eigenvectors $|D(z)\rangle\!\rangle$ of $Z=a\otimes I-I\otimes a^\dag$
with eigenvalue $z$ (this is just a heterodyne
measurement\cite{yuen2}),  now achieving 
$p(z|\alpha)=(\pi\Delta^2)^{-1}\exp[-\Delta^{-2}|z-\alpha|^2]$, with 
variance $\Delta^2=\frac{1-|x|}{1+|x|}$ that, in 
principle, can be decreased at will with the state (\ref{dw})
approaching a state an eigenstate of $Z$ (by increasing the gain of
the downconverter). 

\medskip
{\em Measurement in the presence of noise}.---What happens if the
estimation is performed in the presence of 
noise, namely the channel before and after the unknown transformation
is affected by noise? Here it is instructive to reconsider the problem
of estimating the displacement of a harmonic oscillator in the phase
space in the presence of Gaussian displacement noise, which maps 
states as follows
\begin{eqnarray}
\rho\to\Gamma_{\overline{n}}(\rho)\doteq\int_{\mathbb C}
\frac{\mbox{d}^2\gamma}{\pi \overline{n}}\exp[-|\gamma|^2/\overline{n}]
D(\gamma)\rho D^\dag(\gamma)\;.
\end{eqnarray}
The variance $\overline{n}$ of the noise is usually referred to as
``mean thermal photon number''. The case of Gaussian displacement noise is
particularly simple, since one has the composition law 
$\Gamma_{\overline{n}}\circ\Gamma_{\overline{m}}
=\Gamma_{\overline{n}+\overline{m}}$,
and, moreover $\Gamma_{\overline{n}}[D(\alpha)\rho D^\dag(\alpha)]=
D(\alpha)\Gamma_{\overline{n}}(\rho) D^\dag(\alpha)$. Therefore, if the
measurement is made on the entangled state (\ref{dw}), one can easily
derive a a Gaussian conditional probability distribution with variance
$\delta^2=\Delta^2+2\overline{n}_T$, where $\overline{n}_T$ is the
total Gaussian displacement noise before and after the displacement
$D(\alpha)$, and the noise is doubled since it is supposed equal on
the two entangled Hilbert spaces. On the
other hand, in the measurement scheme with unentangled input (remind
that the optimal is the vacuum), one has
$\delta^2=1+\overline{n}_T$. One concludes that the entangled input is
no longer convenient above one thermal photon $\overline{n}_T=1$ of
noise. This is exactly the threshold of noise above which the
entanglement is totally degraded to a separable state
\cite{simon}, and therefore the quantum capacity of the
noisy channel vanishes\cite{holwern}.

\medskip
{\em Discrimination between two unitaries}.---What about when one
discriminates between only two unitary 
transformations? It is clear that here using an entangled 
input as in Fig. \ref{f-scheme} would be of no benefit, since it is
useless to increase the dimensionality of ${\cal
H}_{out}$. However, we will see that in this case a multipartite 
entanglement would allow {\em perfect discrimination for a finite
number of copies} of the transformation to be determined. 
\par
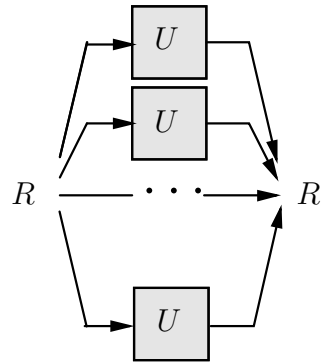
\begin{figure}[hbt]
%\vskip .5truecm
\begin{center}
\setlength{\unitlength}{0.004mm}
\begin{picture}(9540,11898)(0,-10)
\thicklines
\texture{0 0 0 888888 88000000 0 0 80808 8000000 0 0 888888 88000000 0 0 80808 
	8000000 0 0 888888 88000000 0 0 80808 8000000 0 0 888888 88000000 0 0 80808 }
\shade\path(3525,11829)(6000,11829)(6000,9429)
	(3525,9429)(3525,11829)
\path(3525,11829)(6000,11829)(6000,9429)
	(3525,9429)(3525,11829)
\put(4275,10404){\makebox(0,0)[lb]{\smash{\large$U$}}}
\shade\path(3600,2454)(6075,2454)(6075,54)
	(3600,54)(3600,2454)
\path(3600,2454)(6075,2454)(6075,54)
	(3600,54)(3600,2454)
\put(4350,1029){\makebox(0,0)[lb]{\smash{\large$U$}}}
\shade\path(3525,9129)(6000,9129)(6000,6729)
	(3525,6729)(3525,9129)
\path(3525,9129)(6000,9129)(6000,6729)
	(3525,6729)(3525,9129)
\path(2025,7929)(3525,7929)
\blacken\path(2925.000,7779.000)(3525.000,7929.000)(2925.000,8079.000)(2925.000,7779.000)
\path(1125,6129)(2025,7929)
\path(1125,6804)(2025,10629)
\path(1133,6802)(2033,10627)
\path(6000,7929)(7425,7929)(8325,6129)
\blacken\path(7922.508,6598.574)(8325.000,6129.000)(8190.836,6732.738)(7922.508,6598.574)
\path(6000,10629)(7425,10629)(8325,6729)
\blacken\path(8043.926,7279.906)(8325.000,6729.000)(8336.243,7347.364)(8043.926,7279.906)
\path(2025,10554)(3525,10554)
\blacken\path(2925.000,10404.000)(3525.000,10554.000)(2925.000,10704.000)(2925.000,10404.000)
\path(1117,4996)(2017,1171)
\path(1950,1179)(3450,1179)
\blacken\path(2850.000,1029.000)(3450.000,1179.000)(2850.000,1329.000)(2850.000,1029.000)
\path(6150,1179)(7575,1179)(8475,5079)
\blacken\path(8486.243,4460.636)(8475.000,5079.000)(8193.926,4528.094)(8486.243,4460.636)
\path(1125,5529)(3525,5529)
\path(5925,5529)(8325,5529)
\blacken\path(7725.000,5379.000)(8325.000,5529.000)(7725.000,5679.000)(7725.000,5379.000)
\put(4275,7629){\makebox(0,0)[lb]{\smash{\large$U$}}}
\put(-500,5229){\makebox(0,0)[lb]{\smash{\large$R$}}}
\put(3850,5229){\makebox(0,0)[lb]{\smash{\huge$\cdots$}}}
\put(9000,5229){\makebox(0,0)[lb]{\smash{\large$R$}}}
\end{picture}
\end{center}
\caption{When testing between two unitaries $U=U_{1,2}$ it is possible
to achieve perfect discrimination even for nonorthogonal $U_1$ and
$U_2$ for sufficiently large number $N$ of copies of the unitary
transformation, if a $N$-partite entangled state is available for a
measurement scheme as figure (see text).}
\label{f-schemen}\end{figure}

 In an optimized strategy\cite{helstrom} the
minimum error probability in the discrimination of the two output
states $U_1|\psi\rangle$ and $U_2|\psi\rangle$ for any (also
entangled) input state $|\psi\rangle $ is 
\begin{eqnarray}
P_E={1\over2}\left[1-\sqrt{1-4p_1p_2|\langle\psi|U_2^\dag U_1|\psi\rangle|^2}\right]\;,\label{PE}
\end{eqnarray}
$p_1$ and $p_2$ being the {\em a priori} probability of the two
transformations. For simplicity, in the following we set
$p_1=p_2={1\over2}$. Clearly, the optimum input states 
$|\psi\rangle$ are those minimizing the overlap $|\langle\psi|U_2^\dag
U_1|\psi\rangle|$. It is easy to show that the minimum overlap is given by\cite{note1}
\begin{eqnarray}
\min_{||\psi||=1}|\langle\psi|U_2^\dag U_1|\psi\rangle|=r(U_2^\dag
U_1)\;,\label{minoverl}
\end{eqnarray}
where $r(W)$ denotes the distance between the origin of the complex
plane and the poligon whose vertices are the eigenvalues of the
unitary operator $W$.  Moreover, optimizing the overlap over entangled 
$|\psi\rangle$ gives again Eq. (\ref{minoverl}) \cite{note2}. From the
rule of the minimum overlap (\ref{minoverl}) we conclude that the
discrimination is perfect if and only if $z=0\in S(U_2^\dag
U_1)$, namely the poligon of the eigenvalues of $W=U_2^\dag
U_1$ encircles the origin. Then, it is obvious that an entangled
input as in Fig. \ref{f-scheme} would be of no use, since $W$ and
$W\otimes I$ have the same spectrum. However, the situation changes
dramatically if one has $N$ copies of the unitary transformation
$U=U_{1,2}$ to be determined, and a $N$-partite entangled state
is available for a measurement scheme as in
Fig. \ref{f-schemen}. Now the spectrum of $W^{\otimes N}$ must be
considered, and the angular spread $\Delta(W)$ of the eigenvalues is
increased as $\Delta(W^{\otimes N})=\min(N\Delta(W),2\pi)$
($\Delta(W)$ is the angle subtended at the origin by the polygon of
eigenvalues of $W$). Therefore, {\em the discrimination is always exact for
sufficiently many uses} $N$. This result should be compared to the
case of {\em state} discrimination. There, for nonorthogonal states
the probability of failure is always nonvanishing for any $N$.  Here,
instead, for nonorthogonal transformations the
discrimination among unitaries is always exact for $N$ sufficiently
large. It is clear that the above arguments could be extended to the
case of multiple testing,  whenever the strategy leads to an
overlap criterion (as, for example, in Ref. \cite{yuen}). That exact
discrimination between unitaries is virtually possible for
finite number $N$ of uses has been also noticed in Ref.\cite{acin}. 

\medskip
{\em Improving the stability of the measurement}.---In the istances in
which the optimal discrimination between 
transformations is already optimized by a unentangled input, an
entangled state can still be better in achieving a more stable
sensitivity. We have seen that a unentangled input is already optimal
in the discrimination of (one use of) two  unitaries.  A unentangled
input is also optimal in the covariant measurement for abelian
${\mathbb G}$, since the irreducible representations are one
dimensional. Consider, for example, the problem of distinguishing
among displacements on a fixed direction of the phase space, say
$D(x)$, with $x\in{\mathbb R}$. In this case one could use a squeezed
state $|x_0\rangle_s\doteq\exp[{s\over2}((a^\dag)^2-a^2)]D(x_0)|0\rangle$,
with $s>0$, i. e. squeezed in the direction of the ``quadrature''
$X={1\over2}(a^\dag +a)$. Then, a conditional Gaussian probability with
variance $\langle\Delta X^2\rangle={1\over4}e^{-2s}$ is
obtained, which can be narrowed at will by using
$n_s=\sinh^2s$ squeezing photons.  However, if the phase of the
quadrature is slightly mismatched, and the quadrature $X_{\phi}=
{1\over2}(a^\dag e^{i\phi}+ae^{-i\phi})$ is measured instead, then the
variance becomes $\langle\Delta
X_{\phi}^2\rangle={1\over4}(e^{2s}\sin^2\phi+e^{-2s}\cos^2\phi)$, and
the sensitivity is exponentially unstable. Using the entangled input
in Eq. (\ref{dw}), instead, gives the same Gaussian noise
$\Delta^2=\frac{1-|x|}{1+|x|}$, independently on $\phi$, 
by using $n=2|x|^2/(1-|x|^2)$ downconverted photons. 

\medskip
{\em Further generalizations and conclusions}.---Up to now we have
focused our analysis only on discrimination among 
unitaries, however, we could have considered more generally nonunitary
quantum operations, to see that entanglement is still a useful resource
for improving the measurement. For the case of two operations
${\mathrm Q}_1$ and ${\mathrm Q}_2$ the distinguishability is related to the completely
bounded (cb) norm\cite{holwern} $||p_1{\mathrm Q}_1-p_2{\mathrm Q}_2||_{cb}$
which is the supremum over all possible entangled input states of the
trace-distance between the output states. Since the cb-norm is
equivalent to the usual trace-norm for completely positive maps, it
follows that a unentangled state already achieves optimality in the
special case that the difference $p_1{\mathrm Q}_1-p_2{\mathrm Q}_2$
is completely positive.

In conclusion, we have seen that entanglement is a useful resource for
upgrading the quantum measurements which are based on the estimation
of a quantum transformation. It is always of benefit, in improving
either precision or stability. In many cases the measurement precision
becomes in principle unbounded, even when the conventional measurement
is noise limited. The upgrading is  effective in the presence of noise, 
below the threshold of total entanglement degradation.  

\par This work has been cosponsored by the Istituto Nazionale di
Fisica della Materia through the project PAIS-99 TWIN, and by the
Italian Ministero dell'Universit\`a e della Ricerca Scientifica e
Tecnologica (MURST) under the co-sponsored project 1999 {\em Quantum
Information Transmission And Processing: Quantum Teleportation And
Error Correction}. G. M. D. acknowledges support by DARPA grant
F30602-01-2-0528.


\begin{thebibliography}{10}
\bibitem{cocycle} The projective representations are unitary group
representations with an additional phase $\omega(g,h)$ in the
composition law called ``cocycle''. The cocycle must satisfy the
Jacobi associativity constraints, namely that
$\omega(gh,l)\,\omega(g,h)=\omega(g,hl) \,\omega(h,l)$ and
$\omega(g,g^{-1})=\omega(g,e)=1$,  for $g,h,l\in{\mathbf G}$, $e$
being the identity element.
\bibitem{note1} In fact,  by expanding $|\psi\rangle=\sum_j\psi_j|w_j\rangle$  
over the eigenvectors $|w_j\rangle$ of $W=U_2^\dag
U_1$ with eigenvalue
$e^{i\gamma_j}$, consider the convex set ${\cal S}(W)\doteq\{z=\sum_j
|\psi_j|^2 e^{i\gamma_j}\,, ||\psi||^2=1\}$, namely the  polygon whose
vertices are the eigenvalues of $W$. Then, the minimum of 
$|\langle\psi|U_2^\dag U_1|\psi\rangle|$ is just the minimum $|z|$
with $z\in{\cal S}(W)$, namely the distance of the polygon from the
origin of the complex plane.
\bibitem{note2} Using entangled states
$|\psi\rangle\equiv|E\rangle\!\rangle$ corresponds to find the minimum
of $|\mbox{Tr}[W(E^\dag E)^T]|$, which is the average
$\sum_j\rho_j|\langle\lambda_j|U_2^\dag 
U_1|\lambda_j\rangle|$ for $(E^\dag E)^T=\sum_j\rho_j
|\lambda_j\rangle\langle\lambda_j|$, whence the minimum
distance is always obtained for a $E^\dag E$ pure, i. e. for a
unentangled $|\psi\rangle$.
\end{thebibliography}
\end{document}